# Circumstellar Dust Created by Terrestrial Planet Formation in HD 113766

C. M. Lisse[1], C. H. Chen[2], M. C. Wyatt[3], and A. Morlok[4]

[1] Planetary Exploration Group, Space Department, Johns Hopkins University Applied Physics Laboratory, 11100 Johns Hopkins Rd, Laurel, MD 20723   carey.lisse@jhuapl.edu

[2] NOAO, 950 North Cherry Avenue, Tucson, AZ 85719   cchen@noao.edu

[3] Institute of Astronomy, University of Cambridge, Madingley Road, Cambridge CB3 0HA, UK   wyatt@ast.cam.ac.uk

[4] Department of Earth and Planetary Sciences, Faculty of Science, Kobe University, Nada, Kobe 657-8501, Japan   morlok70@kobe-u.ac.jp

Proposed Running Title: **Terrestrial Planet Dust Around HD 113766A**

Please address all future correspondence, reviews, proofs, etc. to:

Dr. Carey M. Lisse

Planetary Exploration Group, Space Department

Johns Hopkins University, Applied Physics Laboratory

11100 Johns Hopkins Rd

Laurel, MD 20723

240-228-0535 (office) / 240-228-8939 (fax)

Carey.Lisse@jhuapl.edu

# ABSTRACT

We present an analysis of the gas-poor circumstellar material in the HD 113766 binary system (F3/F5, ~16Myr), recently observed by the *Spitzer* Space Telescope. For our study we have used the infrared mineralogical model derived from observations of the Deep Impact experiment. We find the dust dominated by warm, fine (~1 um) particles, abundant in Mg-rich olivine, crystalline pyroxenes, amorphous silicates, Fe-rich sulfides, amorphous carbon, and colder water-ice. The warm dust material mix is akin to an inner main belt asteroid of S-type composition. The ~440 K effective temperature of the warm dust implies that the bulk of the observed material is in a narrow belt ~1.8 AU from the 4.4 $L_{solar}$ central source, in the terrestrial planet-forming region and habitable zone of the system (equivalent to 0.9 AU in the solar system). The icy dust lies in 2 belts, located at 4-9 AU and at 30–80 AU. The lower bound of warm dust mass in 0.1 - 20 µm, $dn/da \sim a^{-3.5}$ particles is very large, *at least* $3 \times 10^{20}$ kg, equivalent to a 320 km radius asteroid of 2.5 g cm$^{-3}$ density. Assuming 10m largest particles present, the lower bound of warm dust mass is at least 0.5 $M_{Mars}$ The dust around HD 113766A originates from catastrophic disruption of terrestrial planet embryo(s) and subsequent grinding of the fragments, or from collisions in a young, extremely dense asteroid belt undergoing aggregation. The persistence of the strong IR excess over the last two decades argues for a mechanism to provide replenishment of the circumstellar material on yearly timescales.

# 1. INTRODUCTION

We report here on an analysis of mid-IR observations of HD 113766, a young (~16 Myr old), F3/F5 binary stellar system, located at a distance of 130 pc from the Earth, with two component stars of nearly identical age characterized by early F spectral types in the Sco-Cen association. Attention had been called to this system since its association with a detection in the IRAS Point Source Catalogue (IRAS 13037–4545; Backman & Paresce 1993). More recent work by Meyer *et al.* (2001) and Chen *et al.* (2005, 2006) have confirmed that the system exhibits unobscured photospheres and a large IR excess ($L_{IR}/L_* = 0.015$), but no detectable HI emission (Figure 1). HD 113766A thus belongs to the small, but growing, class of post-T Tauri objects characterized by ages of 5–30 Myr, with large quantities of excess mid-IR emission from hot dust, while lacking in significant gas. Similar objects have been identified in recent work : β Pic (Okamato *et al.* 2004, Telesco *et al.* 2005), η Tel, and HD 172555, members of the β Pic moving group (Zuckerman & Song 2004); η Cha (Mamajek *et al.* 1999) and EF Cha (Rhee *et al.* 2007), members of the η Cha cluster; HD 3003, a member of the Tucana/Horologium moving group (Zuckerman *et al.* 2001); and 8 objects in h- and χ-Persei (Currie *et al.* 2007a, b). It should be noted that the majority of post-T Tauri systems, e.g. the ~10 Myr HD99800B (Furlan *et al.* 2006) and the ~30 Myr HD12039 (Hines *et al.* 2007) of recent note, demonstrate IR excesses on the order of $L_{IR}/L_* = 10^{-4}$, but from relatively cold dust with few strong mid-IR spectral features.

Of these young post-T Tauri objects dominated by warm dust, HD113766 stands out as having the largest relative IR excess and strongest mid-IR spectral features (Rhee *et al.* 2007; Figures 1 & 2). What makes HD113766 so different? Five possibilities come to mind: the strong excess emission (Figure 2) comes from **(i)** residual primordial material, **(ii)** ongoing evaporation of material from a large swarm of primitive icy planetesimals (comets), **(iii)** collisions of many bodies within a massive comet or asteroid (processed, differentiated planetesimals) belt, **(iv)** recent collision between two large comets or asteroids in a planetesimal belt; or **(v)** recent collision between two proto-planets (e.g., the lunar formation event). (Note that for this work, a comet is defined as any < 100 km, ~1/2 volatile ice, low density and strength object with little to no process-ing of its constituent material vs. the material found in the proto-solar nebula, while an asteroid is a < 1000 km body with < 20% volatile content, moderate to high density and strength, which has undergone significant alteration due to thermal and or collisional evolution. A proto-terrestrial planet, derived from numerous asteroidal bodies, is a highly evolved and differentiated, high density and strength body with a > 1000 km.)

The first scenario **(i)** is most likely ruled out from previous work. The disk is relatively devoid of gas, as demonstrated for neutral hydrogen by Chen *et al.* (2006) (and by proxy for other gaseous species) and thus most primordial gaseous material has been removed. The extremely strong observed spectral features are direct evidence for the abundance of small (~1 um) particulates, which have very short in-system lifetimes (on the order of years) due to the effects of radiation pressure ($\beta \sim 1$) and Poynting-Robertson (P-R) drag (Burns *et al.* 1979; Chen *et al.* 2006). This is a very different situation than is found in our stable, mature solar system debris disk (i.e., the interplanetary dust cloud), which is dominated by large particles of size 20–2000 μm (Grogan *et al.* 2001) and dynamical lifetimes of $10^6$–$10^7$ years.

Because the HD113766A phenomenon appears persistent from 1983 (Infrared Astronomical Satellite, or IRAS) until 2005 (*Spitzer*), we find it must be persistent on the timescales of decades, implying a continuous source of dust replenishment. The dynamical constraints, coupled with the estimated age of the system, suggest that the dust is being continuously replenished through sublimation of a large reservoir of planetesimals or collisions in such a reservoir. Thus we must invoke scenarios like **(ii)** and **(iii)** above. A possible, but less likely, scenario is that the dusty material was created in an very recent impulsive fragmentation event involving an asteroidal, cometary, or protoplanetary body, with the dusty material produced slowly (on the order of decades) clearing out of the system (scenarios **(iv)** and **(v)** above). In this case, any gas created by the fragmentation event must have been rapidly evolved, ionized, swept up by the local stellar wind, and removed from the system; otherwise it would have been detected.

The occurrence rate of debris disks containing hot dust around sun-like stars is very low, ~2% (Bryden *et al.* 2006, Chen *et al.* 2006). While recent work suggests that the hot dust of most mature sun-like stars is transiently regenerated, as it is present in quantities far in excess of that ex-pected to arise from sublimation of comets (Beichman *et al.* 2005) or slow collisional grinding of asteroids left over from the era of planet formation (Wyatt *et al.* 2007), HD113766A is different, since its age is too young to consider it as a mature debris disk. Rather it must have emerged only recently (within the last 10 Myr) from its protoplanetary disk phase, and terrestrial planetesimal formation is most likely ongoing (terrestrial planets form on timescales of 10–100 Myr; Wetherill 1990 and references therein; Yin *et al.* 2002; Chambers 2004 and references therein).

During such an epoch large quantities of hot dust may be expected for a number of reasons. Terrestrial planets are thought to grow by accumulation of smaller objects, through the process of collisional accretion (Kenyon & Bromley 2004a, 2006). A few large bodies, or 'oligarchs,' of radii > a few hundred kilometers, emerge from the swarm of initial km-sized objects, and dominate the system. Micron-sized dust is created during this aggregational phase, in quantity on the order of a fraction of an Earth mass. The closer to the system primary the dust is created, the 'drier' (or more volatile poor) and warmer the dust will be. The resulting 'oligarchs' stir up the remaining planetesimals onto eccentric orbits, causing them to collide and fragment, clearing upwards of 90% of the objects while producing a further collisional cascade and more circumstellar dust, in amounts rivaling the mass of the Ceres or even the Moon (Table 3). The oligarchs become planets in this phase, objects like the Moon are created, and much of the collisionally produced dust produced previously is swept up, while radiation pressure, P-R drag, and the stellar wind from the central source remove the rest (Kenyon & Bromley 2004b).

Enhanced warm circumstellar dust produced by the sum total of these terrestrial planet-forming processes, in amounts on the order of an oligarch ($M_{Ceres}$ or larger), should be created and observable in the 10–100 Myr timeframe. As the system relaxes from the era of planet formation over the next few Gyr, the dust density decreases by a few orders of magnitude, but eventually steady-state collisional grinding of small, leftover asteroidal bodies in planetesimal belts support the bulk dust cloud at a low level, and stochastic asteroid fragmentation events in a 1–2 % of stars produces rings of fresh dust analogous to the dust bands observed in the solar system zodiacal light (Bottke *et al.* 2005; Nesvorny *et al.* 2003, 2006) and the debris belt found around HD 69830 (Beichman *et al.* 2005; Lisse *et al.* 2007).

The early studies of Schutz *et al.* (2005), using ESO TIMMI2 spectra, determined that the dusty material lay in a circumstellar belt around one of the HD113766 stars, and suggested that the 10 μm feature is dominated by crystalline silicate (forsterite) and large, amorphous silicates; $SiO_2$, which is correlated with the presence of forsterite in Herbig AeBe ISO spectra, was not detected. Chen *et al.* (2006) revisited this analysis, and determined that HD 113766 possesses a large parent-body mass (0.1 $M_\oplus$ or 260 times that mass in the solar system main asteroid belt) and a high single blackbody grain temperature, $T_{gr}$ = 330 K, suggestive of debris at terrestrial planet temperatures. In a more detailed analysis, using a dust model of crystalline forsterite in addition to amorphous carbon, olivine, and pyroxene, and a single-temperature blackbody continuum to fit the 5.5–35 μm IRS spectrum, they inferred a low crystalline silicate fraction of 4.1% for the dust. From their determination of the presence of a cooler blackbody continuum ($T_{gr}$ = 200 K) in addition to the hot silicate and carbon grains ($T_{gr}$ = 600 K), they suggested that this system possessed two planetesimal belts, analogous to our solar system's asteroid and Kuiper belts. The location of these hypothetical planetesimal belts was in the range $r_*$ = 0.5–2.3 AU from the primary for the warm dust, comparable to the position of the asteroid belt in our own solar system, and at $r_*$ = 4–5 AU for the cold dust.

No allowance was made in either of these spectral models, however, for effects due to particle size or composition. The recent findings on solar system primitive body mineralogy resulting from the Deep Impact and Stardust experiments allow us to greatly improve the models. *Spitzer* IRS observations of the Deep Impact experiment in July 2005 created a new paradigm for understanding the infrared spectroscopy of primitive solar nebular (PSN) material — the observed spectrum of fresh, interior ejected material was the most detailed ever observed. Further, the experiment provided a direct study of the thermal behavior of primitive nebular material in a known radiation field, allowing for empirical determination of the distance of the material in another system if the primary's luminosity is known. Decomposition of the Deep Impact *Spitzer* spectra implied the presence of 7 classes of materials, including silicates, carbonates, phyllosilicates, water-ice, amorphous carbon, PAHs, and metal sulfides (Lisse *et al.* 2006). The consistency of the results with the cometary material returned by the Stardust spacecraft from comet 81P/Wild 2, the *in situ* Halley flyby measurements, and the Deep Impact data return provide a fundamental cross-check for the spectral decomposition models. Further application of the model to the decomposition of the mod-IR spectra of comet C/1995 O1 (Hale-Bopp) and the circumstellar material found around the young stellar object (YSO) HD 100546 proved to be facile (Lisse *et al.* 2007a). In the next study, our group used the DI infrared mineralogical model to study the nature of the dust in the bright, dense debris disk found around the mature (2–10 Gyr old) K0V star HD 69830 (Beichman *et al.* 2005). We were able to show it to be markedly different from any of the cometary systems we had studied: lacking in carbonaceous and ferrous materials but including small icy grains, and appearing much more like the composition of a disrupted, ~30 km radius P- or D-type asteroid from our solar system (Lisse *et al.* 2007b and references therein). The radiative temperature of the dust implied that the bulk of the observed material is at ~1.0 AU from the HD69830 primary.

In this paper, we utilize the *Spitzer* IRS observations of Chen *et al.* (2006) and our infrared mineralogical model to further investigate the nature of the dust found around HD 113766A. We find that the circumstellar dust in the HD 113766A system is processed and differentiated and cannot be derived from cometary material; that it is much more rocky than the olivine-rich material found in the HD 69830 circumstellar belt, dominated by silicates (~75% crystalline), metal sulfides, and amorphous carbon, akin to the material found in S-type asteroids; that the composition of the refractory elements Si, Mg, Fe, Ca and, Al is close to solar, and thus cannot be derived from the crust of a highly differentiated proto-planet; and that it is located in an belt at ~1.8 AU from the primary. Combined with the low estimated age of the system, ~16 Myr (based on its membership in the Lower Centaurus Crux part of the Sco-Cen stellar association), the similarity of the material to the most common asteroid type in the inner (terrestrial) portion of the solar system main asteroid belt, the very large amount of material present (0.5% of the lunar mass, or the mass of an S-type asteroid of radius 320 km), and the strong likelihood that this material is constantly being replenished, we find this result to be compelling evidence for ongoing terrestrial planet formation around HD 113766A. We find that the most likely source for the large amount of material detected by *Spitzer* and IRAS, is the ongoing collisional grinding of an extremely dense, young asteroid belt. It is also possible, but less likely, that the observed dust is due instead to the breakup and complete fragmentation of a large (> 320 km radius) S-type body, on the order of a terrestrial planet embryo size, and as large as the largest asteroid in our solar system. ***Ei-*** *ther of these possibilities are produced by formation processes associated with the building of terrestrial planets.*

## 2. OBSERVATIONS

The first indications of a large IR excess for the HD113766 system were found in the IRAS sky survey (Backman & Paresce 1993; Figure 1). Using mid-IR imaging photometry from the 6.5m Magellan, Meyer *et al.* (2001) found excess emission suggesting the presence of circumstellar dust around one of the two stars in the system, HD113766A (Figure 1), with a range of blackbody temperatures from approximately 290-440 K. (*N.B.* - Only one of the stars, HD113766A, shows a major IR excess, and throughout the text we use the term HD 113766A dust or excess to refer to the circumstellar dust around HD 113766A, and the term HD113766 to refer to the entirety of the binary system, when two component stars are considered as a unit or are unresolved.) The unresolved excess flux observed at 4.8, 11.7, and 18.0 μm appeared to originate from within 20, 50, and 80 AU, respectively, of HD113766A. Including data obtained from the HIPPARCOS/TYCHO photometric database (Figure 1), Meyer *et al.* estimated the reddening and intrinsic stellar luminosity of the two binary components of HD 113766. Calculation of pre-main-sequence stellar evolution trajectories in the luminosity-temperature plane derived a stellar luminosity for HD113766A of 4.4 $L_{solar}$, and an age for the stars of 10–20 Myr, consistent with their kinematic membership in the Lower Centarus Crux sub-group of the Sco-Cen OB association. We adopt both of these values for the work presented here.

Chen *et al.* (2005) targeted 130 F-, G-, and K-type members of the young, nearby Sco-Cen association using *Spitzer* MIPS imaging photometric observations to study the transition from primordial to mature circumstellar disks, and found HD113766 to be one of 14 with appreciable excess 24 and 70 μm flux. In their follow-on study, which provided the data used in this work, Chen *et al.* (2006) obtained *Spitzer* Infrared Spectrograph (IRS) 5-35 μm study of 59 main-sequence stars with previously reported IRAS 60 μm excesses (Sadakane & Nishida 1986; Sylvester *et al.* 1996; Walker & Wolstencroft 1988). HD 113766 was one of 5 conspicuous systems in their survey with large IR excess ($L_{IR}/L_* = 0.015$) and extremely pronounced spectral features vs. the spectral continuum.

For their IRS observations of HD113766, Chen *et al.* utilized a combination of the Short-Low (5.2–14.0 μm, λ/Δλ ~ 90), Short-High (9.9–19.6 μm, λ/Δλ ~ 600), and Long-High (18.7-37.2 μm, λ/Δλ ~ 600) modules. A total of 1719 independent spectral points were obtained over the range 5.2-36 μm. In order to avoid time-consuming peak-up, the observatory was operated in IRS spectral mapping mode where a 2 × 3 raster (spatial × dispersion) centered on the star was performed (Watson *et al.* 2004). The reduction and analysis of the spectra was conducted with the *Spitzer* IRS instrument team's SMART program, V15 (Higdon *et al.* 2004). Relative calibration of the spectral orders, and fringing in the long wavelength data were important issues that had to be dealt with in treating the data. We refer the reader to Chen *et al.* (2006) for more details of the IRS data reduction. The Chen *et al.* (2006) IRS spectrum for HD113766 is overplotted on the stellar photometry in Figure 1.

The 5-35 μm HD113766A disk excess flux studied in this work was calculated by removing the stellar binary photospheric contribution from the IRS spectrum of Chen *et al.* (2006). As HD 113766 is a binary system with a projected separation of 170 AU (1.3"), these objects were not separated by the *Spitzer* IRS and, thus, must be treated as a unit for purposes of photospheric modeling and removal. The photospheric contribution was modeled by assuming that the combined HD 113766A and HD 113766B spectrum is represented by an F3–F5 member of the Lower Centaurus Crux in Sco-Cen (de Zeeuw *et al.* 1999) with an estimated age of 16 Myr (Mamajek *et al.* 2002). Both component stars were assumed to have solar abundances, log g = 4.5, and E(B − V) = 0.01 (determined using the Cardelli *et al.* 1989 extinction law). The stellar photospheric fluxes of the objects were then estimated by minimum $\chi^2$ fitting of published unresolved photometry from the literature to model the combined Kurucz stellar atmospheres for the two sources, using only bandpasses with wavelengths shorter than 3 μm and 2MASS (Cutri *et al.* 2003). The photosphere removed flux is presented and compared to other significant mid-IR dust spectra in Figure 2a.

## 3. MODELS
### 3.1 The Deep Impact Tempel 1 Dust Model
To understand the information derived from the HD 113766A excess IRS spectra, it is important to summarize the Deep Impact experiment and the Tempel 1 Dust Model. Details of the spectral analysis have been described in the literature in the Supplementary Online material for Lisse *et al.* 2006, and in the main text of Lisse *et al.* 2007a. We only list the critical highlights here.

*Spitzer* IRS 5–35 μm spectra were taken immediately before and after the Deep Impact encounter, which occurred when Tempel 1 (T1) was 1.51 AU from the Sun on 2005 July 4. The material that was ejected from the nucleus from depths as large as 30m was pristine and largely unaltered, due to the structural weakness of the material and the low escape velocity (~1 m sec$^{-1}$) from the nucleus (A'Hearn *et al.* 2005). At the same time, it was de-aggregated from loosely-held fractal particles into individual sub-fractal components (A'Hearn *et al.* 2005; Lisse *et al.* 2006; Sunshine *et al.* 2006). The observed material had cooled from effects due to the impact within minutes, and the separation of the ejecta from the ambient coma dust was cleanly made. The resulting highly structured spectrum of the ejecta showed over 16 distinct spectral features at flux levels of a few Janskys (Lisse *et al.* 2006) that persisted for more than 20 hours after the impact. The pre-impact spectrum showed almost no features and was well fit by a blackbody spectral model with temperature near LTE, indicative of the predominance of large, optical thick dust particles in the ambient coma.

***The results of the Deep Impact experiment were observed by Spitzer, as well as 80 other observatories across the spectrum, allowing for multiple independent verifications of the behavior of a known quantity of astrophysical dust in a know radiation field. E.g., Harker et al. (2005), Keller et al. (2005), Sugita et al. (2005), and Schleicher et al. (2006) have all published particle size distributions consistent with our findings, and the Deep Impact near-IR spectrometer measured consistent effective temperatures for the ejecta (A'Hearn et al. 2005). The temperature and particle size distribution of the ejected material did not have to be modeled or assumed; it was measured, as a result of one of the few astrophysical experiments performed to date.***

The emission flux from a collection of dust is given by

$$F_{\lambda, \text{mod}} = \frac{1}{\Delta^2} \sum_i \int_0^\infty B_\lambda(T_i(a,r_*)) Q_{abs,i}(a,\lambda) \pi a^2 \frac{dn_i(r_*)}{da} da$$

where $T$ is the particle temperature for a particle of radius $a$ and composition $i$ at distance $r_*$ from the central star, $\Delta$ is the range from *Spitzer* to the dust, $B_\lambda$ is the blackbody radiance at wavelength $\lambda$, $Q_{abs}$ is the emission efficiency of the particle of composition $i$ at wavelength $\lambda$, $dn/da$ is the differential particle size distribution (PSD) of the emitted dust, and the sum is over all species of material and all sizes of particles for the dust. Our spectral analysis consists of calculating the emission flux for a model collection of dust, and comparing the calculated flux to the observed flux. The emitted flux depends on the composition (location of spectral features), particle size ( feature to continuum contrast), and the particle temperature (relative strength of short vs. long wavelength features), and we discuss each of these effects below.

**Composition.** To determine the mineral composition the observed IR emission is compared with the linear sum of laboratory thermal infrared emission spectra. As-measured emission spectra of randomly oriented, 1 μm-sized powders were utilized to determine $Q_{abs}$ in order to avoid the known inaccuracies and artifacts inherent in mathematical modeling (e.g., Mie theory) of strong emission features. The material spectra were selected by their reported presence in interplanetary dust particles, meteorites, in situ comet measurements, YSOs, and debris disks (Lisse *et al.* 2006). Over 80 different species were originally tested for their presence in the T1 ejecta, and have been checked for their presence in other astrophysical dusty systems. The list of materials tested against the SST spectra included multiple silicates in the olivine and pyroxene class (forsterite, fayalite, clino- and ortho-enstatite, augite, anorthite, bronzite, diopside, and ferrosilite); phyllosilicates (such as saponite, serpentine, smectite, montmorillonite, and chlorite); sulfates (such as gypsum, ferrosulfate, and magnesium sulfate); oxides (including various aluminas, spinels, hibonite, magnetite, and hematite); Mg/Fe sulfides (including pyrrohtite, troilite, pyrite, and niningerite); carbonate minerals (including calcite, aragonite, dolomite, magnesite, and siderite); water-ice, clean and with carbon dioxide, carbon monoxide, methane, and ammonia clathrates; carbon dioxide ice; graphitic and amorphous carbon; and the neutral and ionized PAH emission models of Draine & Li (2007).

The phase space search excluded the vast majority of mineral species from the T1 ejecta. ***From species comprising the best-fit model, Lisse et al. (2006) found convincing evidence for only the following as the majority species in the Tempel 1 ejecta:*** crystalline silicates like forsterite, fayalite, ortho-enstatite, diopside, ferrosilite and amorphous silicates with olivine and pyroxene like composition; phyllosilicates similar to nonerite; sulfides like ningerite and pyrrohtite; carbonates like magnesite and siderite; water gas and ice; amorphous carbon (and potentially native Fe:Ni; and ionized PAHs. These 7 classes of minerals (15 species in all) are the ones we have used in modeling other systems, including HD113766.

Successfully applying our methodology to the spectra of 3 other solar system comets, 3 extra-solar YSOs, and two mature exo-debris disks to laboratory thermal emission spectra of the materials has found additional support for thermal IR emission signatures due to Ca/Fe/Mg-rich silicates, carbonates, phyllosilicates, water-ice, amorphous carbon, ionized PAHs, and Fe/Mg sulfides, in varying proportions. This list of materials compares well by direct comparison to numerous in situ and sample return measurements (e.g., the Halley flybys and the STARDUST sample return; see Lisse *et al.* 2007a for a detailed list), providing a series of strong checks of its validity. A subset of this list was found to fit the HD 113766A spectrum well (Table 1; Figure 3).

Of special note, our ability to detect water-ice and gas has also been checked by comparing the results for the solar system comets versus their location with respect to the ice line. Comet SW-3 (Sitko *et al.* 2008, in preparation) at 1 AU showed only water gas emission; Comets Hale-Bopp (Malfait *et al.* 1998) at 2.8 AU and SW-1 (Stansberry *et al.* 2004) at 5.2 AU showed only water-ice; comet Tempel 1 at 1.5 AU showed a mix, as the rapid excavation of material from the comet by the hypervelocity impact of the DI spacecraft ejected water-ice from the nucleus in a non-equilibrium fashion, faster than the ice could sublimate (Sunshine *et al.* 2007).

**Particle Size Effects.** Particles of 0.1–1000 um are used in fitting the 5–35 um data (although results to date have shown a sensitivity only to the 0.1–20 μm particle size range), with particle size effects on the emissivity assumed to vary as

$$1 - Emissivity(a, \lambda) = [1 - Emissivity(1um, \lambda)]^{(a/1um)}$$

The particle size distribution (PSD) is fit at log steps in radius, i.e., at [0.1, 0.2, 0.5, 1, 2, 5,…100, 200, 500] μm. Particles of the smallest sizes have emission spectra with very sharp features, and little continuum emission; particles of the largest sizes are optically thick, and emit only continuum emission. Both the T1 ejecta spectrum and the HD 113766A spectrum show strong, sharp, high contrast features, indicative of the presence of small (~1 μm) grains. For the Tempel 1 ejecta, the PSD was found to be unusually narrow, consisting predominantly of 0.5–2.0 μm particles (Lisse *et al.* 2006). For the HD 113766A excess, a power law spectrum of particle sizes, including both large and small grains, was found to be necessary to fit the data.

**Particle temperature.** Dust particle temperature is determined at the same log steps in radius used to determine the PSD. Particle temperature is function of particle composition and size at a given astrocentric distance. The highest temperature for the smallest particle of each species is free to vary, and is determined by the best-fit to the data; the largest, optically thick particles (1000 μm) are set to the LTE temperature, and the temperature of particles of in between sizes is interpolated between these extremes by radiative energy balance. We model the unresolved dust excess around HD 113766A as a relatively localized dust torus, and a disk as a superpositional sum of individual torii. It is often difficult to distinguish between the two cases. The material dominating the emission is the hottest and densest material, and for most disks, this is the unobscured material closest to the primary star in the observing beam, as the circumstellar material density and temperature both decrease with distance from the central star. On the other hand, if the circumstellar material is highly extended, the large range of temperatures for the dust serves to smear out any sharp spectral features. This is not the case observed for HD113766.

The T1 ejecta were, by experimental design, all highly localized at 1.51 AU from the Sun. We use this fact to *empirically* determine the effective distance of the emitting material from HD113766A. To do this, the best-fit model temperature for the smallest and hottest (0.1 -1 μm) particles for each material from our analysis is compared to the temperatures found for the (0.5 - 2.0 μm) particles of the Tempel 1 ejecta (Lisse *et al.* 2006), using the relation

$$T_{dust} = T_{T1ejecta}(L_*/L_{solar})^{1/4}(1.51AU/r_*)^{1/2}$$

where $T_{dust}$ is the temperature of the dust around HD113766A, $T_{T1\ ejecta}$ are as given in Lisse *et al.* 2006 (~ 340 K), $L_*$ = bolometric luminosity of HD113766A, and r* is the distance of the dust particle from HD113766A.

**Model Summary.** Our method has limited input assumptions, uses physically plausible emission measures from randomly oriented powders, rather than artificially derived Mie values, and simultaneously minimizes the number of adjustable parameters. The free parameters of the model are the relative abundance of each mineral species, the temperature of the smallest particle of each mineral species, and the value of the particle size distribution at each particle size (Table 1). The total number of free parameters in the model used to fit the IRS HD113766A spectrum = 15 relative abundances + 15 hottest particle temperatures + 1 power law size index = 31. Best-fits are found by a direct search through (composition, temperature, size distribution) phase space.

It is important to note that our methodology has allowed us to get beyond the classical, well known olivine-pyroxene-amorphous carbon composition to the second-order, less emissive species like water, sulfides, PAHs, phyllosilicates, and carbonates. On the other hand, there are limitations to our methods. There is no petrological or isotopic information, and the results returned are for bulk averages of the observed systems. For example, only very abundant species with strong emission features (>10% of the silicate emission peaks) will be detectable. In order to cover the large phase space of possible minerals present, we assume a linear mix of the extreme endstates of the mineral each mineral system, and a linear shift in the band positions and strengths between the endstates. E.g., we linearly adjust the balance of forsterite (Fo100, or $MgSi_2O_4$) and fayalite (Fo100, or $Fe_2Si_2O_4$) to fit the observed spectrum, allowing us to determine the total number of each atom present, but cannot distinguish between the presence of Fo50 ($FeMgSi_2O_4$) and a 50–50 mix of (Fo100 + Fa100). The values in the compositional tables should be interpreted in thusly. We also cannot distinguish easily between "glassy silicate of non-stoichiometric but near olivine (or pyroxene) composition" and "amorphous silicate of olivine (or pyroxene) composition" and so we assume the presence of stoichiometric glasses when modeling the glassy silicates. Emission by relatively cold material, and optical depth effects, can both serve to obscure material in the observed system, so that the amount of material reported in this work is a lower limit to the true system mass.

Despite these limitations, we are able to determine the overall amounts of the different major classes of dust-forming materials (olivines, pyroxenes, sulfides, water, etc.) and the bulk elemental abundances for the most abundant atoms in these materials (C,O, Si, Mg, Fe, S, Ca). Apply-ing our analysis, with a series of strong 'ground truth' checks of its validity, is highly diagnostic for interpreting mid-IR spectra of distant dusty systems like YSOs, debris disks, and PNs.

**3.2 Application of the Tempel 1 Dust Model to the HD113766A IR Excess Observations**

**3.2.1 Qualitative Spectral Comparison.** To zeroth order, marked similarities are found at the gross level between the spectral features for the HD 113766A IR excess and those seen in comets Hale-Bopp and Tempel 1, YSO HD 100546, and the mature asteroidal debris disk system HD 69830, motivating the in-depth spectral analysis presented here (Figure 2a). The quality of the Tempel 1 spectrum, taken for an object at 0.75 AU distance under a controlled astrophysical experiment, is readily apparent. The HD 113766 excess spectrum, while somewhat noisy, was obtained for the most distant object in the set—131 pc away—and is surprisingly bright, indicating a large amount of emitting dust surface area is present in the observing beam.

Most of the obvious differences in the spectra HD 100546 and Hale-Bopp versus the spectra are due to temperature — the dust in Tempel 1, HD 113766, and HD 69830 are much warmer on the average and, thus, have relatively more emission in the 5–10 μm region. Our analysis is done in emissivity space, as identification of emission lines due to different mineral species is easily and quickly done once the gross effects of temperature are removed (compare Figures 2a and 2b). Conversion of the IR excess flux to an emissivity spectrum was performed by dividing the measured fluxes by a best-fit blackbody (with a value of $T_{gr}$ = 490K found from $\chi^2$ minimization testing, somewhat higher than the single grain temperature of 330K, but comfortably between the 200K cold dust and 600K warm dust reported in Chen *et al.* 2006; Figure 2b).

Many of the spectral features are seen at similar levels in all the emissivity spectra, including silicate emission features at 9.8, 11.2, 16.2, 19, 24, and 34 μm. These are the features that dominate the HD113766A excess spectrum. Other major spectral features seen include the very strong PAH emission for HD 100546 at 6.2, 7.7, and 8.6 μm, water gas emission at ~6 μm for HD100546 and the comets, the carbonate peak for Tempel 1 at 6.5–7.5 μm, the amorphous silicates at 8–9 μm in HD 100546, Hale-Bopp, and Tempel 1. The HD 100546 spectrum is dominated in the 5–8 μm region by spectral features due to carbon-rich species like PAHs and carbonates, and water gas. The HD113766A excess shows no evidence of emission from any of these species, suggesting the HD113766A material has been processed to remove these volatile, non-refractory materials (brown oval). On the other hand, unlike HD113766A, HD69830 shows a strong lack of emission due to more easily processed pyroxene at 8–16 μm, and a superabundance of the most refractory silicate olivine. The HD113766A silicaceous material is not as processed as the HD69830 material.

**3.2.2 Mathematical Spectral Analysis.** More quantitatively, the resulting emissivity spectrum (Figure 2b) was compared, using the methodology described in §3.1, to linear combinations of laboratory emission spectra of 15 candidate mineral species (there are no obvious gas emission lines), selected for their reported presence in YSOs, solar system bodies, dusty disk systems, and interplanetary dust particles (IDPs). We present here our best-fit compositional model (Figure 3; Table 1) consisting of the fewest and simplest dust species possible that result in a consistent fit to the observational data. A set of components was tested exhaustively before the addition of a new species was allowed, and only species that reduced the $\chi^2_\nu$ below the 95% confidence limit (C.L.) were kept. It is important to emphasize that while the number of parameters [composition, temperature, particle size distribution (PSD)] may seem large, there are, instead, very few detected species for the 1611 independent spectral points and 8 strong features obtained at high SNR by the SST/IRS over the 5–35 μm range. It was, in fact, extremely difficult to fit the observed spectrum within the 95% C.L. of $\chi^2_\nu = 1.06$. As discussed in §3.1, when performing our spectral reduction, the detailed properties of the emitting dust (i.e., the particle composition, size distribution, and temperature) all have to be addressed. These properties are thus all products of the modeling, available for interpretation, which we present in the next section. The possible range of each of the derived parameters was found by determining all models with $\chi^2_\nu < 1.06$. The resulting conservative ranges (2σ) of the abundance of each of the types of mineral species are ± 20% (Table 1), of the maximum temperatures of the smallest dust are ± 20 °K, and of the slope of the PSDs are ± 0.2.

## 4. DERIVED PROPERTIES OF THE DUST

**4.1    Composition and Atomic Abundance.** The derived compositional abundances from our best-fit dust model, with $\chi^2_\nu = 1.03$, are given in Table 1. As a test of the robustness of our best-fit model, the last column in Table 1 also gives the best-fitting $\chi^2_\nu$ found after deleting a particular species from the compositional mix and re-fitting the data. Extensive searching of the possible compositional mixes, size distributions, and particle temperatures was conducted. The phase space structure was found to be simple, with one deep minimum centered at the best-fit solution.

Overall, what we found is a very simple composition unlike the comets and comet-dominated YSO systems we have previously studied. The HD113766A circumstellar dust spectrum can be reproduced using a simple mix of Mg-rich olivines (amorphous and crystalline), crystalline pyroxenes, and metal sulfides. The disk material appears to lack carbonaceous species like carbonates and PAHs, and to have no Fe-rich olivines. The olivine crystalline fraction is 81%, similar to that found for the primitive comets and cometary exo-systems (HD 100546, HD 163296), but the pyroxene crystalline fraction is also very high, ~74%, unlike cometary material, and the overall amount of amorphous pyroxene is highly depleted, suggesting that there has been some destruction of the less refractory amorphous pyroxene component in the HD 113766A material. These crystallinity values are sensible given the extremely pronounced (vs. the continuum) silicate features, but are in sharp contrast to the value of 4.1% derived by Chen et al (2006). The compositional makeup is almost equally divided, by mole, amongst the silicate, metal sulfide, and amorphous carbon species.

We also find a detection of warm water-ice at 10-30 um in our spectra (Figure 4). The relative molar amount of water-ice, ~0.17, is at the high end of the range found for the asteroidal material detected in the HD 69830 system (0.10–0.18). While thus plausible from a compositional point of view, it is not clear from a temperature analysis if the water-ice is intrinsic to the hot refractory dust debris belt we find at 1.8 AU, nor if water-ice would have lifetimes long enough to be found co-located with the warm dust belt. No significant emission due to water gas at 5–7 um was found, consistent with the low neutral hydrogen/dust ratio found by Chen *et al.* (2006) for the system. (If water gas was present in large quantities, it would undergo UV photolysis to produce OH + H within 1 day in the HD113766A radiation field.) The lack of water gas emission also argues against a large sublimation rate and the rapid creation of fresh, new water-ice to replace the sublimation losses.

Our derived relative atomic abundances for the HD113766A circumstellar material, H:C:O:Si:Mg:Fe:S:Ca:Al = 1.2 : 0.78 : 4.3 : 1.0 : 0.83 : 1.0 : 0.79 : 0.053 : 0.080 (assuming Si =1.0) are near solar for all the refractory elements Si:Mg:Fe:Ca:Al, and are super-solar for S (Table 2). The estimated H, C, and O abundances are similar to those found for the comet-like systems (Figure 5), and are reasonable for the amount of CHON material typically captured into a small primitive body of limited gravitational influence, with the remainder residing in the gas reservoir of the nebula and not the solids, in species like $H_2$, $H_2O$, $CO_2$, $CO_2$, $CH_4$, etc. (Lisse *et al.* 2007a). (A similar level of C depletion is also found in CI chondrites and the Earth's crust ; Robnov & Yaroshevsky 1969; Loders 2003; Jura 2006).) The metallicity of the HD 113766 system has been estimated as close to solar, Fe/H = –0.02 (Nordstrom *et al.* 2004). As the abundances of the common refractory elements Si:Mg:Fe:S:Al were found to be very close to solar for comet and comet-dominated YSO material (Lisse *et al.* 2007a; Figure 5), and for CI chondrites and the Earth's crust, this strongly suggests that the HD 113766A circumstellar dust is similarly derived from a well sampled mix of primordial nebular material, and little atomic differentiation has occurred.

**4.2 Temperature and Dust Location Versus the Primary.** Our best models for the unresolved dust excess around HD 113766A are for localized dust torii. No extended runs of emitting material were required to fit the data, and as such we do not consider here the more complicated, higher degree of freedom disk models. The toroidal models are motivated by the narrow dust structures found in many of the Hubble Space Telescope (HST) images of debris disks (Kalas *et al.* 2005, 2006), as well as the single-temperature distributions found for many of the objects studied by Beichman *et al.* (2006) and Chen *et al.* (2006). E.g., the material detected around HD 69830 was determined to reside in a torus at 1 ± 0.2 AU from the K0V primary. (Recent observations using the Gemini 8m telescope are consistent, having confirmed that there is no obvious material from outside 2 AU (i.e., no extended disk) and probably demonstrated a marginal detection of the torus at 1 AU; Beichman *et al.*, in preparation.) Further support for a narrow spatial distribution of the dust is found in the sharp, narrow spectral emission features (Figure 3). If the warm dust were spread out over many AU, the dust grains would have a wide range of temperatures depending on particle size and heliocentric distance, and the ensemble sum of these different temperature grains would produce a broad, continuum-like emission, such as is found for the optically thin but highly extended disk of Beta Pic, (Chen et al. (2007)).

The 'apparent' continuum dust temperature for the HD 113766A excess, $T_{gr}$ = 490 K, is derived by directly matching a blackbody radiance curve to the observed run of flux with wavelength, and is much higher than temperatures derived for the typical post-T Tauri system of similar age (e.g. $T_{gr}$ = 150K for the ~10 Myr HD99800B (Furlan et al. 2006) and $T_{gr}$ = 110K for the ~30 Myr HD12349 (Hines et al. 2007), or the $T_{gr}$ = 250 K found for the roughly coeval YSO Be9V system HD 100546 (Lisse et al. 2007a). But it is near to the $T_{gr}$ = 440 K found for the warm asteroidal dust found around the 2–10 Gyr old HD 69830 K0V primary at a distance of 1 AU (Lisse et al. 2007b). Using our more detailed modeling methodology, we find that our best-fit temperature for the **smallest** dust particles of each material (0.1–1.0 µm), which superheat significantly above Local Thermal Equilibrium (LTE), is 490 K for the amorphous carbon, ~450 K for the olivines, and ~420 K for the pyroxenes. These values are all comfortably close to each other, as expected physically, and follow the rough $(Q_{abs}(optical)/Q_{abs}(IR))^{0.25}$ law (Lisse et al. 1998), with the most optically absorbing material, amorphous carbon, the warmest, and the most optically transparent pyroxenes somewhat cooler than the olivines, similar to the spectral modeling results found by Wooden et al. (1999) for comet Hale-Bopp. As for other systems, we note that the 'apparent' continuum dust temperature, $T_{gr}$, found by finding the best-fit single blackbody match to the entire IRS spectrum, is close to the amorphous carbon hottest grain temperature, which dominates the continuum emission at short wavelengths (Figure 3). As we have shown in § 4.1, though, amorphous carbon is only one of a number of species in present in the system, and thus the single blackbody best for temperature should not be over interpreted other than to be used as a coarse measure of the overall dust temperature and location. I.e., given $T_{gr}$ = 490 K, we can quickly ascertain that the average refractory dust particle temperature is between 400-500 K; and that given a 4.4 $L_{solar}$ primary, the emitting dust must reside at a few AU from the central source.

Scaling empirically from the Deep Impact results (§ 3.1), we find, for a 4.4 $L_{solar}$ HD113766A (Meyer et al. 2001; Chen et al. 2005) that we are observing a warm belt of dust located in the inner part of the system at 1.8 ± 0.10 (1σ) AU (Figure 6). If this dust were in the solar system, undergoing irradiation at 1 $L_{solar}$, and had the same effective temperature range for its dust (450

K for the smallest, warmest particles and 290K for large, optically thick blackbody particles), it would lie at ~0.90 AU, very close to where the Earth lies. (Note that we have made these calculations ignoring the effects of the second star on the net insolation of the dust. Such an approach is physically valid, since the effect of the second star is negligible on dust at 1.8 AU from HD113766A. If the two stars were of equal luminosity, and HD113766B was ~170 AU away, then the effect of the second star would be to increase the equilibrium temperature by a factor of $[1+0.25(1.8/170)^2] = 1.00003$).

***This location of the warm HD113766A dust in the heart of the terrestrial habitability zone of the system immediately suggests that we can expect similar nebular chemistry for the material that formed the Earth and the warm material detected in the HD113766A disk, modulo the effects of the higher photospheric temperature (e.g., increased UV flux), and the potentially more massive protoplanetary disk mass in the HD113766A system.***

The apparent effective temperature of the smallest and warmest water-ice particles in our best-fit model to the IRS data, ~200 K, is the temperature for ice at or just below sublimation equilibrium with vacuum; the temperature of pure water-ice cannot vary above this value, as increases in input energy merely vaporize more water molecules. However, only very pure water-ice will be unabsorptive and have long-term lifetimes; as little as a few percent by mass of dark, absorbing material will cause a substantial increase of the radiative energy deposition into the ice, an increase in its temperature, a strong change in the dependence of temperature on particles size, and a drastic increase in its sublimation rate and reduce particle lifetimes to less than a few days (Lien 1990). If the water-ice were intimately mixed together with the hot silicate/metal sulfide/amorphous carbon dust also detected by *Spitzer*, it would be actively subliming and require constant replenishment, as the lifetime for dirty water-ice material at 1.8 AU from a $L_* = 4.4\ L_{solar}$ primary is on the order of minutes to hours (Lien 1990). Since there is no indication of substantial water gas in the system, we conclude that the water-ice detected is either extremely pure, co-located with the hot circumstellar dusty material at ~1.8 AU, but poorly decoupled from the primary's radiation field, or, much more likely, the water-ice is 'dirty'; i.e. mixed with as little as a few % of dark, absorbing material, but is at distances from the primary large enough so that sublimation losses are very small; i.e. the material lies outside the system's "ice line". The ice line for the smallest particles (0.1 µm) in our calculation lies at 9 AU (assuming $L_* = 4.4\ L_{solar}$); for the largest (1000 µm), at 4 AU. The smallest particles will reach a temperature of 200K at any distance inside of 9 AU from HD 113766A, but will evaporate quicker the closer they come to the star. Thus we cannot be more definite abut the location of the ice, due to the uncertainty brought into the physical picture by energy losses due to sublimation. We can say that we do not see significant emission due to water gas, so sublimation losses must be low, and it is most likely the water-ice lies in the 4-9 AU region, and the two sources of emission are simply confused along the line of sight in the large *Spitzer* IRS beam.

More support for the existence of unassociated icy dust reservoirs comes from studying the HD113766A system at wavelengths longer than measured by the IRS. The *Spitzer* Multiband Imaging Photometer (MIPS) 70 µm channel is especially sensitive to cold (30 - 100K) dust, and water-ice has a strong emission feature at 65 µm. A MIPS photometric flux of 350 +/- 35 (1σ) mJy at 70 µm was reported for the system by Chen *et al.* (2006). Extrapolating our best-fit model spectrum out to longer wavelengths (Figure 4), we find a predicted flux of at most 70 mJy at 70 µm. Something other than the hot dust at 1.8 AU must be providing the bulk of the observed 70 µm emission. A simple calculation shows that a second reservoir of very cold water-ice (T ≤ 75K) or large blackbody particles (T ≤ 75K) can produce the remainder of the observed 70 um flux, while adding negligibly to the 5–35 µm emission (Figure 4). We cannot distinguish between these two models for the very cold dust from the given data, nor can we determine whether or not the dust is colder than 75 K. We do note that dust at 75 K would be at 30 - 65 AU from the HD 113766A primary, depending on particle size and composition, and thus resident in the system's equivalent of a Kuiper Belt; dust much farther out than 80 AU would be dynamically unstable versus HD 113776B, at 170 AU distance from HD 113766A (Figure 6).

Thus our model results are consistent with the conclusion that there is a an icy dust population at a distance of ~9 AU of HD 113766A (equivalent to icy material in the giant planet region of the solar system), and a second population of colder icy Kuiper Belt dust farther from the primary at r = 30 to 80 AU (T ≤ 75K), contributing IR emission only at the longest wavelengths, (Figures 4 and 6). Abundant cold icy dust is highly plausible; systems with 70 um *Spitzer* excesses due to icy dust at distances ~100 AU from the primary are relatively common (Bryden *et al.* 2006), and there are a now a few examples of MS systems as young as HD113766 with copious amounts of distant, cold dust (e.g. TWA 7, Matthews *et al.* 2007; HD 99800B (Furlan *et al.* 2006); and HD 12349 (Hines *et al.* 2007)).

**4.3 Size Distribution and Total Mass.** The best-fit dust particle size distribution (PSD) found from our modeling, $dn/da \sim a^{-3.5\pm0.2}$ for particles with radius $0.1 < a < 20$ µm, argues for dust dominated by small particles in its emitting surface area, and for a dust mass dominated by particles of the largest sizes. The predominance of small particle surface area is why we see the strong emission features—the dominant particles are optically thin, and the feature-to-continuum ratio is high. A similar PSD was found for Fomalhaut's disk (Wyatt & Dent 2002), and is the size distribution that is generally assumed for modeling debris disks (e.g., Augereau *et al.* 2001). A system in collisional equilibrium typically demonstrates a PSD $\sim a^{-3.5}$ (Dohnanyi 1969; Williams & Wetherill 1994; Durda & Dermott 1997). For "real" systems, a size distribution even steeper than $a^{-3.5}$ at small sizes is expected in a collisional cascade, both because of its truncation at the blow-out limit (Thebault *et al.* 2003) and because of the dependence of particle strength on size (O'Brien & Greenberg 2003). The fact that we see an $a^{-3.5}$ PSD in the system argues that the dust we are observing is relatively "fresh," at least on the radiation pressure timescales of years and P-R timescales of decades to centuries for 1 µm dust particles. By contrast, the mature solar system IPD cloud, close to equilibrium with the solar radiation field, has a very flat PSD devoid of small particles, $dn/da \sim a^{-1.4}$ (Grogan *et al.* 2001). Integrating the mass implied by the best-fit PSD, we find a dust mass of $3 \times 10^{20}$ kg detected by the *Spitzer* 5–35 µm IRS spectrum (i.e., mass in particles of 0.1–20 µm in size that contribute appreciably to the observed emission, and thus the $\chi^2_n$ value of the model fit). $3 \times 10^{20}$ kg is ~0.5 % of the lunar mass, or the mass of an S-type asteroid of radius $\geq 320$ km, $10^2$ times the dust mass that is found in the mature dust clouds surrounding HD 69830 and $10^4$ times the dust mass that is found around the Sun (Table 3). We quote this number here as the amount of mass definitively detected by the *Spitzer* observations.

However, the majority of the mass is in the largest, optically thick particles if they are present ($M_{total} \sim a^{0.51}$), and there is no reason to believe that the maximum particle size is ~20µm. While the *Spitzer* IRS measurements are not very sensitive to larger grains, the SED compiled by Chen *et al.* (2006) for HD 113766 includes a high SNR photometric detection of the system at 70 µm (*Spitzer*) and a 3σ upper limit at 100 μm (IRAS). Direct interpolation of our best-fit compositional model, using a dn/da ~ $a^{-3.5}$ PSD for particles of sizes from 0.1 to 100 μm (here we explicitly include up to 100 μm grains in our modeling, as the longer wavelengths being considered have appreciable contributions from larger grains sizes than 20 μm) can only account for at most 20% of the 70 and 100 μm emission, and there must be another source of infrared flux in the system, such as cold dust of T ≤ 75 K (Figure 4). Because it is also possible that all the 70 and 100 μm flux arises from the cold dust, we cannot state conclusively that the warm dust has particles large than 20 μm. Future sub-mm and radio measurements, sensitive to larger particles due to their longer wavelengths, will be required to do so.

If instead, following Beichman *et al.* (2005), we estimate the amount of warm dust mass by assuming a PSD of constant slope up to a maximum size of 10m in the HD113766A circumstellar disk, we find a total mass of at least 3.4 x $10^{23}$ kg, or 5.7 x $10^{-2}$ $M_{Earth}$, or 4.9 $M_{Moon}$, or 0.5 $M_{Mars}$ (Table 3). 3.4 x $10^{23}$ kg is similar to the amount of mass thought to have been present in the nascent solar system asteroid belt (Chambers 2004). Assuming that the emitting dust is derived from collisional grinding in a young, dense asteroid belt (scenario **iv**), we can use another method of mass estimation to give a similar result. Comparing our best estimate of the 0.1-1000 μm HD113766A dust mass to the dust mass estimated to be in the solar system zodiacal cloud today (consisting mainly of 0.1-1000 μm dust particles), we find the HD113766A dust to be ~$10^5$ times as massive. Assuming the collisionally produced dust production rate to go as the total planetesimal belt mass[2], estimating one-half of the solar system zodiacal cloud to be asteroidal in origin, and ignoring small factors on the order of unity, we derive a mass for the HD113766A disk, including all bodies, to be sqrt($10^5$) * $Mass_{asteroid\ belt}$ = 0.15 $M_{Earth}$ = 1.5 $M_{Mars}$. This is consistent with the 10m maximum size estimate derived above, suggesting that there is probably much more dust mass present than definitively detected by the 5-35 μm *Spitzer* measurements. We thus include the 3.4 x $10^{23}$ kg estimate as an upper limit for the range of HD113766 dust masses listed in Table 3.

The dust surface area ~ $a^{-0.51}$, and is dominated by the smallest particles. From direct integration of the best-fit model PSD, we find a total estimated 0.1–20 μm surface area from the IRS data of 1.2 x $10^{22}$ $m^2$, or 0.54 $AU^2$. Assuming the dust is in a ring centered at 1.8 AU from the central star, in an annulus 0.4 AU wide (± 2σ in the best-fit model radius) and with total surface area 1.1 x $10^{23}$ m$^2$, we find a spatial filling factor of 0.11 for the dust. It is interesting to note that the fraction of the illuminated sphere at 1.8 AU covered by dust is $1.2/(4\pi*1.8^2) = 0.029$, close to the $L_{IR}/L_*$ = 0.015 value (Chen et al. 2006). Allowing for energy losses due to scattering (i.e. an average Bond albedo for the dust of 10 - 20%, as found for solar system comets emitting fine dust (Lisse et al. 2002)) improves the agreement even further. We can also say something about the opening angle of the disk : the inclinations of disk particles must be greater than 3.7º (0.064 radians) for the dust to intercept enough of the HD 113766A luminosity to explain the infrared luminosity. Even at inclinations of ~4º, the disk would have to be optically thick along the mid-plane. An optically thick disk also implies that the total disk mass estimate of 3.4 x $10^{20}$ kg presented above are conservative lower limits for the actual total disk mass. The same arguments apply in the case of a 10m largest particle.

## 5. DISCUSSION: NATURE OF THE HD 113766 DUST

In § 1, we presented 5 possible physical scenarios for the source of the warm dust in HD113766: **(i)** primordial material, **(ii)** from ongoing evaporation of material from a large swarm of primitive icy planetesimals (comets), **(iii)** from a collisions of many bodies within a massive comet or asteroid (processed, differentiated planetesimals) belt, **(iv)** from a recent collision between two large comets or asteroids in a planetesimal belt; or **(v)** from a recent collision between two protoplanets. Here we utilize the results of the modeling presented in § 4 to eliminate all incompatible scenarios, then discuss what the source(s) of the dust must be.

**5.1    Non-cometary composition.** The circumstellar material around HD 113766 does not appear to be primitive; i.e., derived from a cometary source, containing abundant water and carbon-bearing species (PAHS, $CO_2$, carbonates, organic ices, etc.), like the circumstellar material around the HD 100546 system or in comets Tempel 1 and Hale-Bopp (Lisse et al. 2006, 2007a). We can verify this hypothesis by directly comparing the HD 113766 and HD 100546 disk spectra (Figure 2), noting the absence of features in the 5–9 μm region due to PAHs, amorphous carbon, water gas, and carbonate species in HD 113766. The HD 113766 material appears instead to be compositionally close to that of igneous rocks on the Earth, lacking in all but the most refrac-tory amorphous carbonaceous component (akin to terrestrial soot or volcanic ash).

We also see evidence for processing of the HD113766A dust in the compositional olivine/pyroxene trend plot of Figure 7, created by analyzing the ISO and *Spitzer* mid-IR spectra of 9 dusty systems using our compostional model (Lisse *et al.* 2007c), along with a ground-truth point calculated utilizing the latest bulk ratios for the Stardust comet Wild2 sample return (Zolensky *et al.* 2007). As the extremes of the trend plot are found in the dust around YSOs and the dust found in the elderly white dwarf systems, the overall trend is for the pyroxene content of a system to decrease with increasing system age and material processing. We attribute the apparent trend to the fact that pyroxene materials are less refractory and thermodynamically stable than olivine materials and, thus, less resistant to high temperature and pressure changes inflicted on system material due to collisions, gravitational accretion, differentiation, radioactive decay, and stellar heating on Myr to Gyr timescales. The relative effect of each of these processes depends not only on the age of the dust, but also on its location with respect to the system primary, and the parent body history of the dust. E.g., compare the SW1 location to that of the 4 solar system comets Tempel 1, Hale-Bopp, SW-3, and Wild 2 in the silicate trend plot of Figure 7. SW1, while also a relatively primitive, icy outer solar system body, is large enough to have undergone significant alteration due to $Al^{26}$ radiogenic heating and solar insolation during its formation (Merk & Prialnik 2006); Toth & Lisse 2006], while retaining a solar atomic abundance (as determined from our compositional modeling).

On the olivine/pyroxene trendline, we find that HD 113766 appears to be more similar to the asteroidal debris-disk material found around HD 69830 (Beichman *et al.* 2005, Lisse *et al.* 2007b) and, by direct comparison, to asteroidal material in the HD 69830 system. It also lies in a similar location as the material emitted from the large, differentiated, refractory solar abundance Centaur body SW1. We conclude from the mineralogical evidence that the HD113766A circumstellar material is derived from a processed and differentiated parent asteroidal/protoplanetary body or bodies, and we can rule out any of the primitive cometary body scenarios **(i)** residual primordial material; **(ii)** for sublimation from comets; **(iii)** for collisional fragmentation from comets in a dense belt; or **(iv)** the disruption of a super-comet planetesimal as the source of the warm dust in HD113766. We are left to consider scenarios **(iii)** and **(iv)** for asteroids or protoplanets, or sce-nario **(v)** the spallation of surface material from a protoplanet during a lunar formation event.

**5.2  Dust Parent Body.** Our derived mineralogy suggest the likely progenitor(s) were similar in composition to an S-type asteroid (in the Tholen Taxonomy, e.g. Lodders and Fegley 1998, Table 13.7m p. 250, and references therein), bodies commonly thought to be the source of much of the terrestrial planets' mass. As discussed in § 1, at 10–20 Myr in our solar system, the giant planets have already formed, nebular gas has mostly cleared, and the terrestrial planets are beginning to coalesce. It is thus very reasonable for us to be observing the effects of dusty aggregation of rocky terrestrial planet material in the *Spitzer* spectra. These effects could be causing dust either by **(iii)** collisional grinding between asteroids in a dense belt as small rocky planetesimals slowly aggregate into larger bodies, or **(iv)** by catastrophic collisions between two large asteroidal bodies, of the size of planetary embryos or oligarchs (~$M_{Ceres}$).

The near-solar atomic abundance of the refractory species in the best fit spectral model cannot distinguish between these two scenarios. Solar abundance dust can be evidence for production from a large collection of differentiated objects, averaging out to solar values; or for the youth and primitive nature of the parent body(s), reflecting the original mix of solar abundance materials from which they formed. We thus made an attempt was made to learn more about the putative parent bodies of the dust observed in the HD11376 system using the mineralogical, as well as the atomic, information derived from the Spitzer spectra, by comparing the best-fit spectral model results for the classes of material present to compositional measurements of solar system meteorites. Following Stoeckelmann and Reimold (1989), a mixing calculation was performed using modal data (normalized by volume or Vol%) of the mineralogical composition of meteorites from the literature (Jarosewich 1990; Papike 1998; Hutchison 2004, Warren 2006). We note that our calculations are highly simplified - only the mineral components occurring in the HD113766A spectrum were used for the calculations. All crystalline olivines were added and used as total olivine for the calculations (same with pyroxene class minerals). All various sulfides were added to sulfides, and all different phyllosilicates were lumped together. All carbon (graphite) was counted as amorphous carbon. For Carbonaceous Chondrites, the carbon contained in volatile species was assumed to be destroyed by any collisional mechanisms and was ignored. For the Iron meteorites, a simplified 'model iron meteorite' with only carbon, iron and sulfide was used.

Allowing for the simplicity of our modeling, the fact that modal data for all meteorite types is not available, and that meteoritical materials show a very wide range in composition, we nevertheless find that three of the mineral class results from HD113766A are highly diagnostic, and allow us to limit the number of potential meteoritic components:

- **Metal Sulfides**: Most meteorites rarely have more than 10 Vol% of sulfides (stony enstatite chondrites can get up to about 16 Vol%). To get to the 25 Vol% of HD113766A, only iron meteorites (highly differentiated meteorites consisting of nickel-iron metal and sulfides, and related siderophile species) are possible. Irons can contain up to about 50 Vol% sulfides.

- **Amorphous Carbon**: Most meteorites contain hardly any carbon. Exceptions are the Irons, Ureilites (relatively primitive achondrites, or stony meteorites) and Carbonaceous Chondrites (very primitive meteorites, of near-solar abundance; hereafter CC). However, in most of the CC carbon is in organics, for which no evidence is seen in the IRS spectra (and probably would not survive a collision). Ureilites contain about 14 Vol% carbon, mostly graphitic. In Iron meteorites, clasts of graphite are also common in abundances similar to that in HD113766A.

- **Phyllosilicates**: Phyllosilicates as alteration products usually occur in the primitive CC meteorites like Murchison CM2, Orgueil CI1 or Tagish Lake C2.

For HD113766, these constraints imply that all allowable meteorite fits ($\chi^2 < 1$) show a mixture of stony meteorite (achondrite, specifically Ureilite) at 37-59 Vol%, sulfide-rich iron meteorite at 27-33 Vol%, and primitive CC between 7.9 and 8.3 Vol%. Stony meteorites dominate the mix - the best fits were found for a mixture of 37% Ureilite, 27% Iron, 8% Murchison and 27% Ordinary LL3.4 chondrite (another stony meteorite). However, as long as the three species (Ureilite iron, and CC) were in the mixture, one could add a mix of most other stony meteorite types (at 20 - 30 Vol%) and still get a good fit. This finding is consistent with an S-type asteroid as the parent body for the observed dust.

How can we interpret these results? The linkage between specific types and classes of meteorites to the asteroids is sometimes difficult. There are problems in observing the mineralogical composition of an asteroid surface in a quality good enough to distinguish different types of meteorites, and in allowing for the effects of space weathering on the spectral properties of asteroid surface material. Classification is based on few spectral characteristics (optical/near-IR) and mineral ratios, which can indicate various groups of meteorites. It is also important to take into account that no meteorite probably represents the composition of a whole asteroid - it is just the only sample we have, and sample biases are large. E.g., current models predict that multiple large protoplanetary bodies, long since collisionally disrupted or dynamically removed, were formed from the protoplanetary disk in the early asteroid belt, and upwards of 99% of the original mass has been removed in forming the present day asteroid belt. In the protoplanetary disk, Fe and S are of similar atomic abundance to Si and Mg, and the Fe has to wind up somewhere as aggregation occurs. So for any stony material linked to an achondrite or S-type asteroid, it is reasonable to expect in a primitive, unevolved circumstellar disk that there should also be material similar to what is found in Iron meteoritic material - contained in metal/metal sulfide-rich inclusions, or in hidden cores, or in metallic asteroids (e.g., M-types) (Burbine *et al.* 2002). Yet metallic asteroids are and meteorites are relatively rare in the solar system.

Given these caveats, complete disruption of a mildly processed, semi-differentiated, Ureilite-dominated stony parent body, with metallic/metal-sulfide inclusions and/or core, is the best candidate we find for producing the HD113766A circumstellar disk material. Ureilites are common primitive achondrites (stony meteorites), the second most abundant after the HED-group. The Ureilite parent body (S-type) is thought to be > 100 km in radius (Warren *et al.* 2006), on the same size scale as the body we estimate to have produced the HD113766A dust. Carbonaceous Chondrites, the most common and most primitive of the meteorites formed from the protoplanetary disk, are the likely starting material for Ureilites (Warren *et al.* 2006; Goodrich *et al.* 2007). The melt produced by heating up CCs to get Ureilites is metal sulfide rich, providing the material for the Iron meteoritic component we find in our meteorite model. The CC material required by the meteorite fit could be interpreted as non-altered leftover from the precursor material used to form the Ureilite-dominated stony parent body. Alternatively, the CC material found in our meteoritic fit could have been part of a second, primitive body involved in the two-body collision that created the observed dust. Either physical picture is consistent with the location of HD113766 in the olivine/pyroxene silicate trend line (Figure 7) - the dusty material is more processed than the stuff composing comets, but is less re-worked and differentiated than material found in mature asteroids. Either case supports an asteroidal parent body fragmentation scenario **(iv)** as the source of the observed dust.

It is important to note that we have not allowed here for the effect of collisional shock effects on the initial parent mineralogy in scenario **(iv)**. While the overall effect of shock effects in large body collisions is relatively unexplored and uncertain, and further work needs to be done in this area, we can get an estimate of these effects using the studies on shocked Murchison samples (Tomioka *et al.* 2007). The major effect was the production of abundant amorphous silicate phases, of similar nature to the amorphous silicates used in our spectral modeling. Since these phases are relatively rare in Ureilites, Iron meteorites, and CC, in the impact scenario between two bodies of these types the amorphous materials could be impact products, and evidence for impact produced processes. We also note that there is no reason that the dust has to be created in just one large two-body collision - as in scenario **(iii)**, collisional grinding dust production in a dense asteroid belt - but we reserve judgment on the results of the meteorite modeling concerning this scenario, as the lack of compositional data for all meteorite types at the time of this writing negates our ability to build an entire asteroid belt and average over the expected collision products.

Another possible scenario is that we are witnessing scenario **(v)** - thermal emission from dust created by a massive collision and partial spallation event between two large, highly differentiated proto-planetary bodies, such as is thought to have formed the Earth's Moon (Benz *et al.* 1986; Canup 2004). Similar arguments were put forth by Telesco *et al.* (2005), who found a large excess "clump" of emission (at ~50 AU) in mid-IR thermal images of the similarly aged Beta Pic system (~12 Myr), and determined that small dust grains making up the "clump" most likely originate from a recent planetesimal collision during the latter stages of outer icy planet formation in that system. However, the material spun off the accreting Earth by the impact of an approximately Mars-sized body was rich in light Mg-rich silicates and Al-oxides (outer crustal material), and poor in Fe and other heavier refractories, unlike the near-solar atomic refractory abundances, and the large molar fraction of metal sulfides, we find for the HD 113766 circumstellar material (§4.1). We can thus rule out a lunar-formation type scenario **(v)** as creating the observed circumstellar material, and focus instead on the interactions of bodies smaller than the Moon and more primitive (less differentiated).

**5.3  Dust location Versus the primary.** The location of the dust excess is in the inner system region, exactly where terrestrial planet formation is expected to occur. The net temperature of the dust is such that the majority of icy materials should be in gaseous form. Little evidence for gas emission has been found in the system, however, allowing us to conclude that the dust-to-gas ratio of the ~440K circumstellar material is very high, favoring rocky planet formation and scenarios **(iii)** or **(iv)** for processed planetesimals (asteroids) as the source of the HD113766A dust. The 1.8 AU distance of the dust belt from the primary is equivalent to a distance of 0.9 AU in the solar system. The potential of material located so close to HD113766A to form a terrestrial planet is not significantly affected by HD113766B at 170 AU average distance (Quintana *et al.* 2007). The location of water in the form of icy dust, at 4-9 AU, apparently straddling the system's snow line, is also quite reasonable and has an analogue in current models of early solar system formation (Kuchner 2003; Raymond *et al.* 2004).

**5.4  Dust Mass.** The mass of the observed warm dust, $\geq 3 \times 10^{20}$ kg , and its equivalent single asterodial body (density = 2.5 g cm$^{-3}$) size, $\geq 320$ km (radius), is the size of the bodies believed to be critical to the formation of Earth-like planets in the latter part of their growth period. If, instead, $3 \times 10^{20}$ kg consisted of cometary material, it would represent the mass of about 1 million average solar system comets (Table 3). However, if derived from a million comet-like objects, there would be a large amount of gas in the system, including water gas, as typical solar system comet ratios of emitted gas to dust mass lie in the range 0.3-3.0 (Lisse *et al.* 2002). We know the HD113766 system is gas-poor, by about a factor of 4, versus the ISM gas/dust ratio of ~100 from the observations of neutral hydrogen lines by Chen *et al.* (2006). While the neutral hydrogen result tells us that much of the proto-stellar nebular gas has been removed, and giant planet forma-tion has finished, it is not very restrictive on the presence of comets in the system. A stronger limit is found by the non-detection of water gas emission in the spectrum analyzed in this work. Using a conservative upper limit, at the 95% C.L. of 0.01 relative surface area (Table 1), we find roughly 1/7th as much water gas to dust as was seen in the comet Tempel 1 ejecta (Lisse *et al.* 2006). The ejecta was shown to have a gas-to-dust ratio of < 0.77, so the HD113766 gas to dust ratio is < 0.1, very low for a cometary system, again disfavoring a source for the dust from cometary emission (scenario **(ii)**).

We can rule out the other possible icy parent body scenario as well. If instead the observed material had been aggregated into a single cometary parent body **(scenario (iv))** before its release in a catastrophic breakup, $10^{20}$ kg of material would have densified and differentiated (Toth & Lisse 2006; Jewitt *et al.* 2007). This processing would transform the original cometary material into substances more typical of the outer icy moons and dwarf icy planets—high density ice phases, large particles of conglomerate rock, aqueous alteration products like carbonates, sulfates, and phyllosilicates—which we do not find dominating the warm dust spectra. On the other hand, this mechanism could have led to the production of the icy dust we see located at ~9 AU and at 30–80 AU from HD 113766A. (Unfortunately, because of the low temperatures for the icy dust, the short wavelength diagnostic features in the IRS passband are highly attenuated. Since we have so little information on the amount and kind of cold dust in these reservoirs, we do not examine them further, but note that future observations with high spatial may very well produce important new findings on the cold dust.)

**5.5   Aged Dust?** Could the system be a very slowly aging or slightly modified primordial disk, as predicted by scenario (i)? Probably not. The free gas-to-dust ratio we find for the circumstellar material is on the order of 1000 times lower than ISM values. Using the ~10 Myr old HD 100546 as an example of a primordial dusty disk system, we note that the HD113766 particle size distribution is much more heavily weighted to larger particles. Whether this is due to evolution of the HD113766 dust due to aggregation and agglomeration creating larger dust, or due to radiation pressure and P-R drag preferentially removing the smallest dust, is not clear. Dynamical effects are certainly important for the ~16 Myr old HD113766 system—for the smallest dust detected, 0.1 μm, the ratio of the radiation force to stellar gravity, $\beta \sim 1.0$, and such dust dust is removed from the system by radiation pressure on orbital timescales as soon as it is created. For the largest warm dust we consider, 20 μm, β ~ 0.03, it would take $10^4$-$10^5$ years for this dust to spiral in by P-R drag (Burns *et al.* 1979), where it would evaporate. As noted in § 5.1, the mineralogy of the warm HD113766 dust is very different from the primitive, comet-like material found around HD100546, being much more processed and evolved. Finally, we note that the amount of dust mass is about a factor of 30 lower than the lower mass limit for HD 100546 from ISO (Table 3), indicating that the large majority of the HD 113766 disk has been sequestered or removed.

In sum, the HD 113766 warm circumstellar material appears to be in a fundamentally non-primordial regime, one much more likely to be derived from collisions in a massive asteroid belt (scenario **(iii)**), or from disruption of a large processed parent body (scenario **(iv)**) (cf. the HD 69830 system, Beichman *et al.* 2005; Lisse *et al.* 2007b) than to be derived from a primordial nebula dominated by primitive cometary material, like HD 100546. Despite the reported lack of strong spectral features on whole-systems scales (Pantin *et al.* 1999; Chen *et al.* 2007), the ~12 Myr old A5 β Pic system may be a closer behavioral analog, as suggested by recent high spatial resolution studies using the Subaru telescope providing evidence for 3 narrow circumstellar belts of enhanced small silicate particles (Okamato *et al.* 2004).

**5.6** **Future work.** Future measurements of this system will be highly valuable in further elucidating the mechanisms currently at play. The two most likely source mechanisms for the dust given here (scenarios **(iii)** and **(iv)**) should have different temporal signatures. The ongoing collisional grinding of an extremely dense, young asteroid belt undergoing collisional aggregation should yield roughly constant spectral behavior with time, perhaps with stochastic increases when fresh collisions occur, while the breakup of a large, > 320 km radius, S-type body should demonstrate a monotonically decaying dust spectral signature and density with time. The relative constancy of the mid-IR excess emission detected for this system (Figure 1) over the 20 years' time between the IRAS and *Spitzer* observations would seem to argue against any impulsive single events and for continual collision processes, although we caution that 20 years may not have been long enough for the dust clearing mechanisms to operate fully, and allow us to distinguish between the two possibilities.

Observations of the system at submm, mm, and radio wavelengths (Herschel, ALMA) will also be highly useful, allowing us to put better lower limits on the total mass of dust and ice in the system. Spatial studies of HD 113766 might be able to distinguish between the different mechanisms by searching for "clumps," or localized condensations of dust formed in a single breakup or collision, versus the smooth distribution expected from collisional grinding throughout a thick asteroid belt. However, to achieve this will require resolution at the 2 AU/120pc *(1"/AU) = 16 mas level, making imaging of the dust torus estimated from this work problematic even with the James Webb Space Telescope (JWST), although useful upper limits to the radius of the torus will be obtained.

## 6. CONCLUSIONS

•We have obtained excellent fits to the 5–35 μm mid-IR spectra of the dusty-disk system HD 113766 using the Deep-Impact-T1 ejecta model.

•Only olivines, pyroxenes, Fe-rich sulfides, amorphous carbon (or native Fe:Ni), and water-ice are found in abundance in the disk material of the system. Apart from the water-ice, the mix of materials is close to that found for common S-type asteroids in our solar system.

•Assuming the central star to have a luminosity $L_* = 4.4\ L_{solar}$, we locate the refractory material causing the observed emission at 1.8 AU. If this material had formed in our solar system, it would have been located 0.9 AU from the nascent Sun. The majority of the detected water-ice may be co-located or as far out as 9 AU from HD 113766. There is also evidence for another zone of cold dust, at 30–80 AU from the system primary.

•The amount of mass in 0.1 μm to 20 um particles responsible for the observed IR emission excess is at least $3 \times 10^{20}$ kg, equivalent to an asteroidal body of 320 km radius (assuming 2.5 g cm$^{-3}$ density). ). Extrapolating up to a 10m largest particle, we estimate the amount of mass present to be at least $3 \times 10^{20}$ kg, or 0.5 $M_{Mars}$.

•Despite a similarity in ages, the ~16 Myr old F3 HD 113766A system is clearly not an older analog of the primitive, giant-planet-forming, > 10 Myr old HD 100546 A0eV system, which is dominated by cometary material and has an extensive, massive primordial disk. The ~12 Myr old A5 β Pic system is a closer analog for solar system formation.

•We find the sources for the observed excess emission to be either the complete disruption of an approximately S-type, ≥ 320 km radius (terrestrial-planet-forming) body or a very dense asteroid belt made up of a large number of small S-type bodies undergoing continual aggregational collisions. Either of these possibilities is a predicted outcome of terrestrial-planet-formation processes, making this system an exciting and potentially groundbreaking object for future study.

## 7. ACKNOWLEDGEMENTS

This paper was based on observations taken with the NASA *Spitzer* Space Telescope, operated by JPL/CalTech. C. M. Lisse gratefully acknowledges support for performing the modeling described herein from JPL contract 1274485 and the APL Janney Fellowship program. The authors would also like to thank M. Meyer for many valuable discussions and comments in improving this work.

## 9. TABLES

### Table 1. Composition of the Best-Fit Model[a] to the SST IRS HD113766A Spectrum

| Species | Weighted[b] Surface Area | Density (g cm$^{-3}$) | M.W. | $N_{moles}$[c] (relative) | Model $T_{max}$[d] (°K) | Model $\chi^2_\nu$ if not included |
|---|---|---|---|---|---|---|
| **Detections** | | | | | | |
| *Olivines* | | | | | | |
| Amorph Olivine (MgFeSiO$_4$) | 0.09 | 3.6 | 172 | 0.19 | 450 | 5.59 |
| ForsteriteKoike (Mg$_2$SiO$_4$) | 0.22 | 3.2 | 140 | 0.50 | 450 | 10.4 |
| Forsterte38 (Mg$_2$SiO$_4$)[e] | 0.14 | 3.2 | 140 | 0.32 | 450 | 6.31 |
| *Pyroxenes* | | | | | | |
| Amorph Pyroxene (MgFeSi$_2$O$_6$) | 0.06 | 3.5 | 232 | 0.09 | 420 | 4.22 |
| FerroSilite (Fe$_2$Si$_2$O$_6$) | 0.07 | 4.0 | 264 | 0.10 | 420 | 2.67 |
| Diopside (CaMgSi$_2$O$_6$) | 0.06 | 3.3 | 216 | 0.09 | 420 | 2.25 |
| OrthoEnstatite (Mg$_2$Si$_2$O$_6$) | 0.04 | 3.2 | 200 | 0.06 | 420 | 1.63 |
| *Phyllosilicates* | | | | | | |
| Smectite Notronite Na$_{0.33}$Fe$_2$(Si,Al)$_4$O$_{10}$(OH)$_2$ * 3H$_2$O | 0.08 | 2.3 | 496 | 0.03 | 450 | 3.75 |
| *Metal Sulfides* | | | | | | |
| Ningerite (Mg$_{10}$Fe$_{90}$S)[f] | 0.26 | 4.5 | 84 | 1.4 | 450 | 12.5 |
| *Organics* | | | | | | |
| Amorph Carbon (C) | 0.06 | 2.5 | 12 | 1.3 | 490 | 8.35 |
| *Water* | | | | | | |
| Water-ice (H$_2$O) | 0.15 | 1.0 | 18 | 0.83 | 200 | 5.56 |
| **Upper Limits and Non-Detections** | | | | | | |
| *Water* | | | | | | |
| Water Gas (H$_2$O) | 0.00 | 1.0 | 18 | ≤ 0.00 | 200 | 1.03 |
| *Carbonates* | | | | | | |
| Magnesite (MgCO$_3$) | 0.00 | 3.1 | 84 | ≤ 0.00 | 450 | 1.03 |
| Siderite (FeCO$_3$) | 0.00 | 3.9 | 116 | ≤ 0.00 | 450 | 1.03 |
| *PAHs* | | | | | | |
| PAH (C$_{10}$H$_{14}$) | 0.02 | 1.0 | <178> | ≤ 0.011 | N/A | 1.03 |

(a) - Best-fit model $\chi^2_\nu$ = 1.03 with power law particle size distribution dn/da ~ a$^{-3.48}$, 5 - 35 µm range of fit.
(b) - Weight of the emissivity spectrum of each dust species required to match the HD113766A emissivity spectrum.
(c) - $N_{moles}$(i) ~ Density(i)/Molecular Weight(i) * Normalized Surface Area (i). Errors are ± 10% (1σ).
(d) - All temperatures are ±10K (1σ).
(e) - Not found in cometary systems to date.
(f) - A ningerite composition of Mg$_{25}$Fe$_{75}$S may fit the data better.

**Table 2. Refractory Dust Atomic Abundances for Solar for Objects Observed by ISO/*Spitzer*[a]**

| Object | H | C | O | Si[b] | Mg | Fe | S | Ca | Al |
|---|---|---|---|---|---|---|---|---|---|
| Tempel 1 ejecta | 3.8e-4 | 0.052 | 0.46 | 1.0 | 0.82 | 0.79 | 0.61 | 0.84 | 1.0 |
| Hale-Bopp coma | 5.0e-05 | 0.13 | 0.23 | 1.0 | 1.1 | 0.97 | 0.63 | 0.45 | 1.4 |
| HD 100546 (Be9V) | 1.1e-4 | 0.35 | 0.26 | 1.0 | 0.88 | 0.76 | 0.63 | 0.00 | 1.2 |
| HD 69830 | 8.8e-06 | 0.088 | 0.18 | 1.0 | 1.1 | 0.43 | 0.00 | 0.93 | 0.0 |
| HD 113766 (F3/F5) | 3.0e-05 | 0.076 | 0.18 | 1.0 | 0.77 | 1.09 | 1.73 | 0.82 | 0.96 |

[a] - Abundance estimates have 2σ errors of ± 20%.
[b] - All abundances are with respect to solar, with the Si abundance assumed to be = 1.0 for normalization purposes.

**Table 3. Derived Total Masses (in beam) for the Objects Observed by ISO/*Spitzer* and Selected Relevant Solar System Objects**

| Object | Observer Distance[1] (pc/AU) | Mean Temp[2] (K) | Equiv Radius[3] (km) | 19 um Flux[4] (Jy) | Approximate Mass[5] (kg) |
|---|---|---|---|---|---|
| Earth | --- | 282 | 6380 | | $6 \times 10^{24}$ |
| Mars | 1.5 | 228 | 3400 | | $6 \times 10^{23}$ |
| Moon | 0.0026 | 282 | 1740 | | $7 \times 10^{22}$ |
| HD 100546 (Be9V) | 103.4 pc | 250/135 | ≥ 910 | 203 | $\geq 1 \times 10^{22}$ |
| HD 113766 (F3/F5) | 130.9 pc | 440 | 300 - 3000 | 1.85 | $3 \times 10^{20} - 3 \times 10^{23}$ |
| Asteroid Belt | 0.1 - 5.0 AU | Variable | | | $3 \times 10^{21}$ |
| HD 69830 | 12.6 pc | 340 | 30 - 60 | 0.11 | $3 \times 10^{17} - 2 \times 10^{18}$ |
| Zody Cloud | 0.1 - 4.0 AU | 260 | | | $4 \times 10^{16}$ |
| Asteroid | 0.1 - 5.0 AU | Variable | 1 - 500 | | $10^{13} - 10^{21}$ |
| Comet nucleus | 0.1 - 10 AU | Variable | 0.1-50 | | $10^{12} - 10^{15}$ |
| Hale-Bopp coma | 3.0 AU | 200 | | 144 | $2 \times 10^{9}$ |
| Tempel 1 ejecta | 1.51 AU | 340 | | 3.8 | $1 \times 10^{6}$ |

(1) - Distance from Observer to Object.
(2) - Mean temperature of thermally emitting surface.
(3) - Equivalent radius of solid body of 2.5 g cm$^{-3}$.
(4) - System or disk averaged flux.
(5) - Lower limits are conservative, assuming maximum particle size determined by *Spitzer* or ISO (20μm), ignoring optical thickness effects. Upper limits assume a maximum particle size of 10m radius.

## 10. -  FIGURES

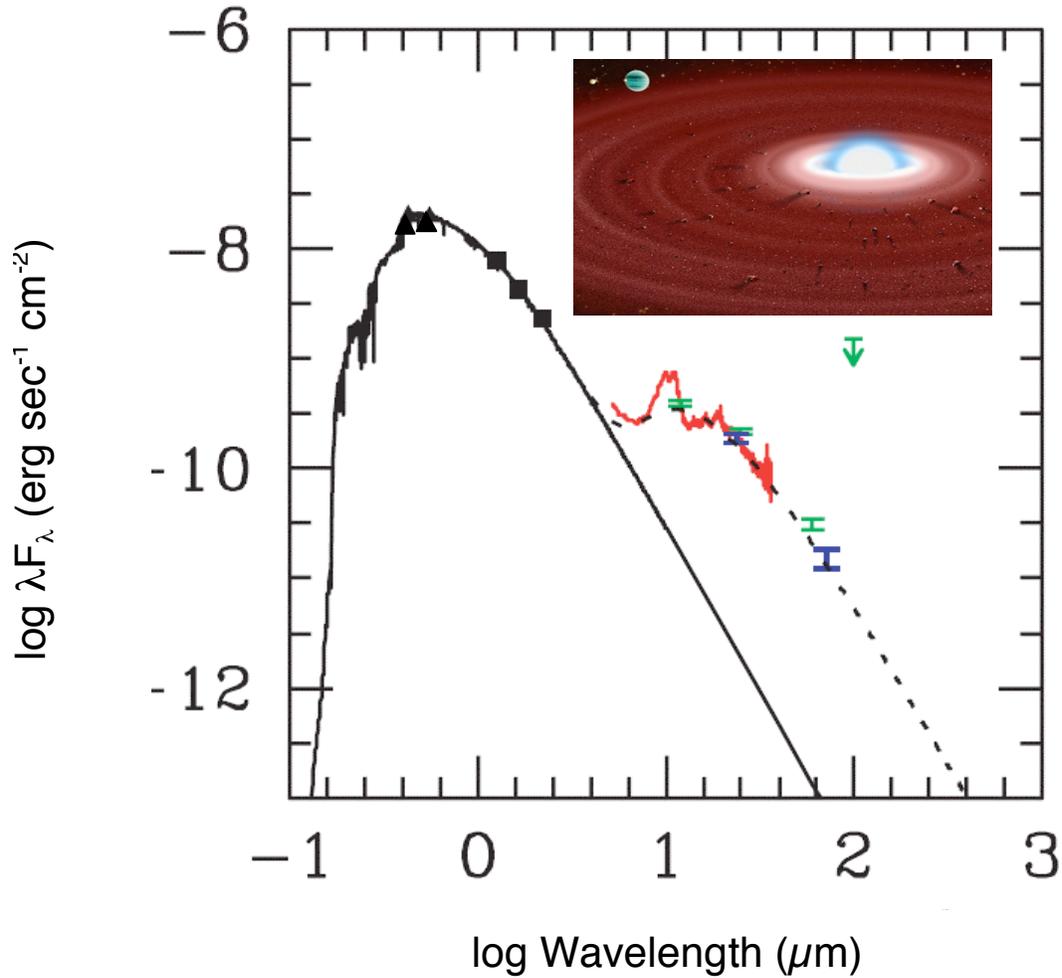

**Figure 1 — *S*pectral energy distribution for HD 113766**, after (Chen *et al.* 2005, 2006). TYCHO (triangles) and 2MASS (squares) photometric measurements are shown with filled black symbols; IRS spectra are shown in red; MIPS photometry, where available are shown with blue error bars; IRAS photometry are shown with green error bars (12–60 μm) and an upper limit arrow (100 μm). The combined photospheric models are shown with a solid black line. The dashed line is the best-fit constant emissivity model to the photometric points.

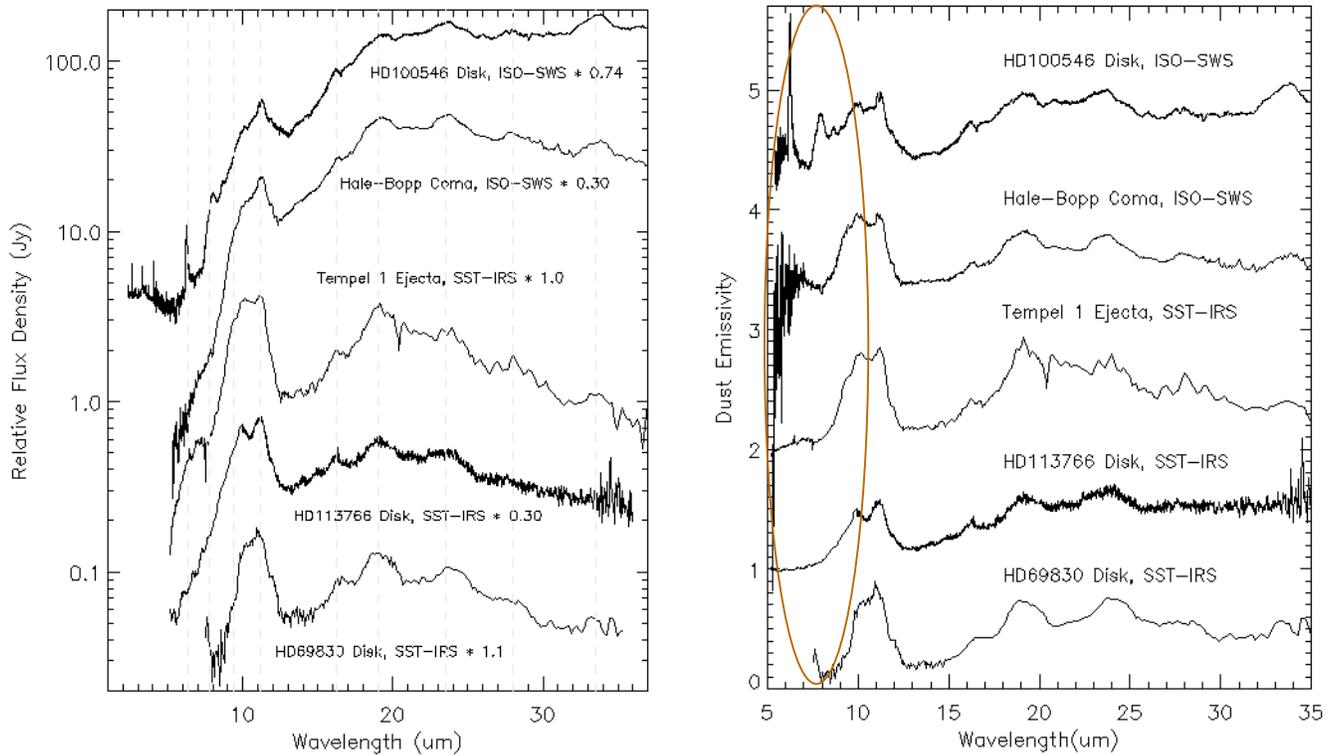

**Figure 2 — (a) Mid-IR spectra and (b) temperature corrected dust emissivity spectra of the extremely young stellar object HD 100546 (after Malfait *et al.* 1998; Lisse *et al.* 2007a), the young terrestrial-planet-building system HD 113766, the mature solar-system-like HD 69830 system (after Beichman *et al.* 2005; Lisse *et al.* 2007b), and the comets C/1995 O1 (Hale-Bopp) and 9P/Tempel 1 (after Lisse *et al.* 2006, 2007a)**, showing the gross similarities and differences (brown oval) in the flux and particle emissivity for the sources. The uncertainty of each measurement can be estimated from the high frequency variations in the data on wavelength scales of 0.05 µm.

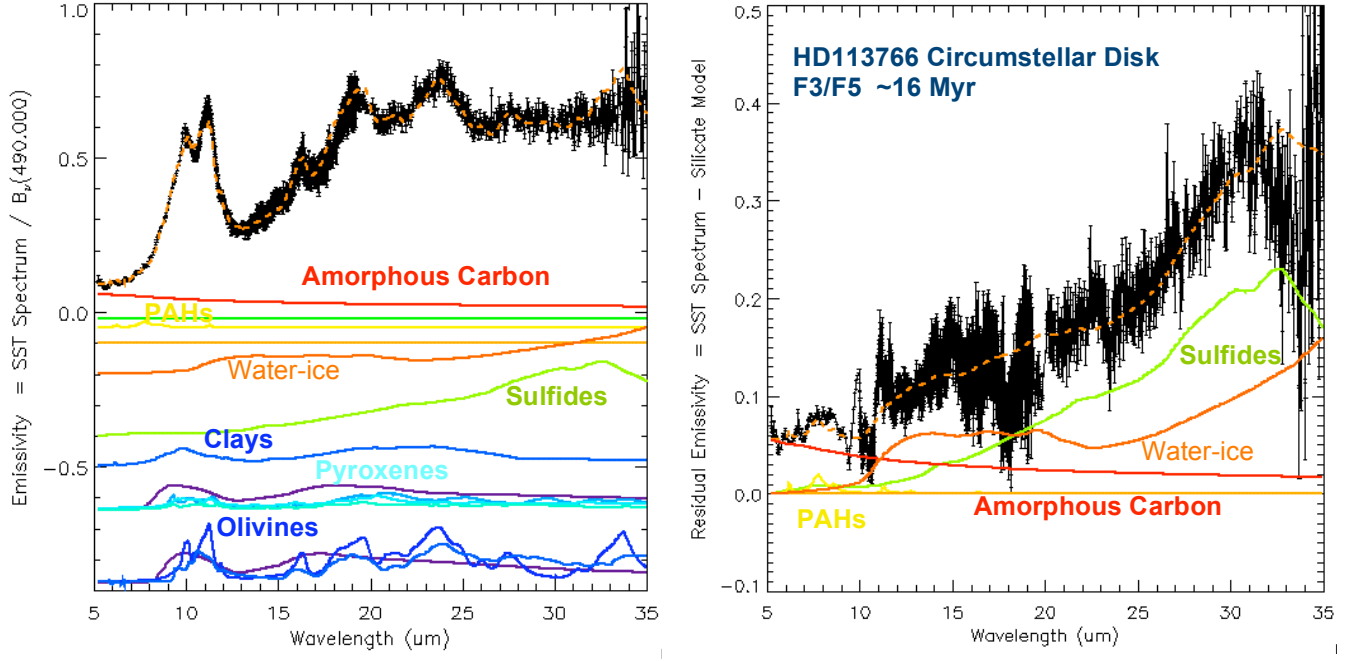

**Figure 3 — (a) *Spitzer* IRS emissivity spectrum of HD 113766, with best-fit spectral decomposition.** The central source's photospheric contribution has been removed using a Kurucz model with a 6870 K photospheric temperature. Error bars are ±2σ. The amplitude of each colored curve denotes the relative amount of that species present in the best-fit model (Table 1). For species with no statistically detectable emission, the curve is a flat horizontal line. Black: SST excess spectrum, divided by a 490 K blackbody. Orange dashed line: best-fit model spectrum. Colored curves: emission spectra for the constituent species, scaled by the ratio $B_\lambda(T_{dust}(a)_i)/B_\lambda(T_{gr})$, with $T_{gr}$ = 490 K. Purples—amorphous silicates of pyroxene or olivine composition. Light blues—crystalline pyroxenes: ferrosilite, diopside, and orthoenstatite. Dark blues—crystalline olivine forsterites. Red—amorphous carbon. Deep orange - water ice. Light orange - water gas. Yellow- PAHs. Bright greens…carbonates: siderite and magnesite. Olive green - ferromagnesian sulfide $Fe_{0.9}Mg_{0.1}S$. The brown oval highlights the 5–9 µm wavelength region where the HD 113766 spectrum differs the most markedly from the HD 100546 SED. (b) Residual emissivity of the circumstellar dust, after the emission due to the dominant silicates has been fit and removed. The remnant is dominated by emission from amorphous carbon, water ice, and metal sulfides. (Note that the $Fe_{0.9}Mg_{0.1}S$ emission spectrum used in our modeling could be improved to obtain a better fit; a ningerite with slightly increased Mg fraction, such as $Fe_{0.75}Mg_{0.25}S$, or a pyrrhotite is most likely indicated, but a good laboratory spectrum of these materials has yet to be measured.) While there are interesting hints of potential water gas emission at 6 µm, and of carbon dioxide gas emission at 15 µm, neither of these features is statistically significant.

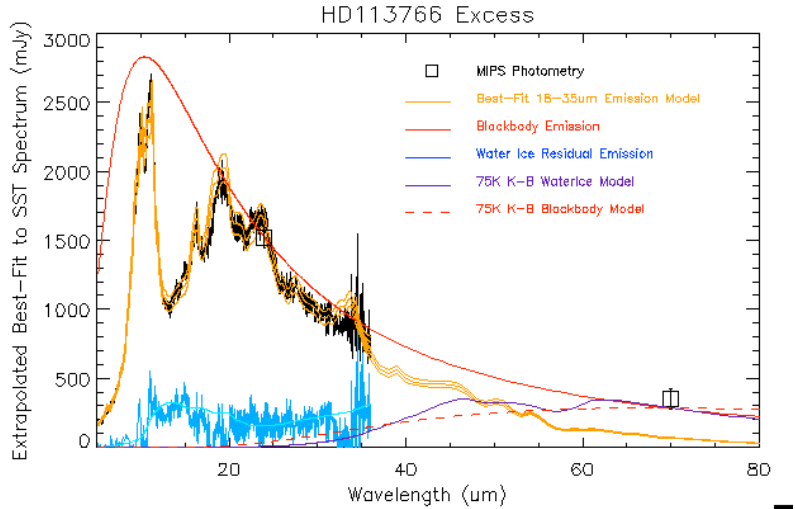

**Figure 4 — Comparison of the *Spitzer* MIPS 70 um photometry and of the best-fit model to the IRS 5–35 um HD 113766 excess.** The MIPS error bars are 2σ. Black - IRS data. Orange—IRS emission model, extrapolated out to 80 μm. The model has been normalized at the nominal, +2σ, and -2σ MIPS 24 um flux levels. The predicted flux of 67 ± 3 mJy from the best-fit model is less than 20% of the measured MIPS 70 μm flux of 350 ± 70 mJy. Red solid line—Warm blackbody model normalized to the 24 and 70 um MIPS fluxes, demonstrating that a single system consisting of large, dark radiators cannot produce both the observed 5–35 and 70 μm emission. Dark Blue—Residual emission due to warm water ice, produced by subtracting all other model fluxes from the IRS data. Light blue line—Warm ice emission model flux. The dark purple curve shows the predicted flux from a reservoir of 75 K water ice particles with PSD dn/da ~ $a^{-3.5}$, required in addition to the warm dust dominating the 5–35 μm spectrum to produce the observed 70 μm MIPS flux. Emission from this population would be undetectable in the IRS bandpass. The dashed red curve shows the predicted flux from a collection of blackbody emitters at T = 75 K, which would also be undetectable in the IRS bandpass. It would be very hard to distinguish between these two models.

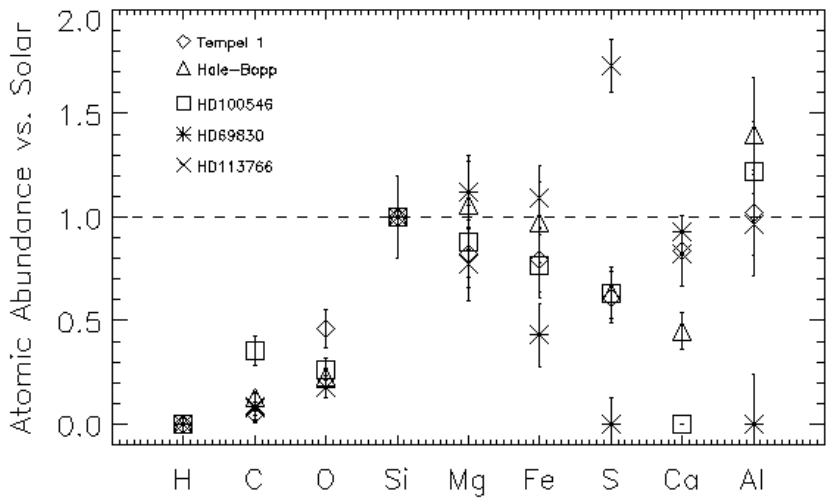

**Figure 5 — Atomic abundances for the refractory dust around HD 113766,** compared to the published values for the comets Tempel 1 and Hale-Bopp, the comet-dominated YSO HD100546, and the asteroidal debris disks system HD 69830 (Lisse *et al.* 2007a,b). Abundances are given vs. solar, assuming Si = 1.0 Error bars for the relative measures are ± 20% (2σ). Diamonds = Tempel 1, Squares = HD 100546, Triangles = Hale-Bopp, Stars = HD 69830. The nominal solar value is denoted by the dashed line. HD 113766 is normal vs. solar in the refractory species Si, Mg, Fe, Ca, and Al, similar to the cometary systems, and appears markedly abundant in S. Allowing for the large reservoir of unretained C, H, and O in volatile species, this dust appears to consist of a representative mix of the HD 113766 primordial nebular materials. Little atomic differentiation has occurred, except for possibly for S.

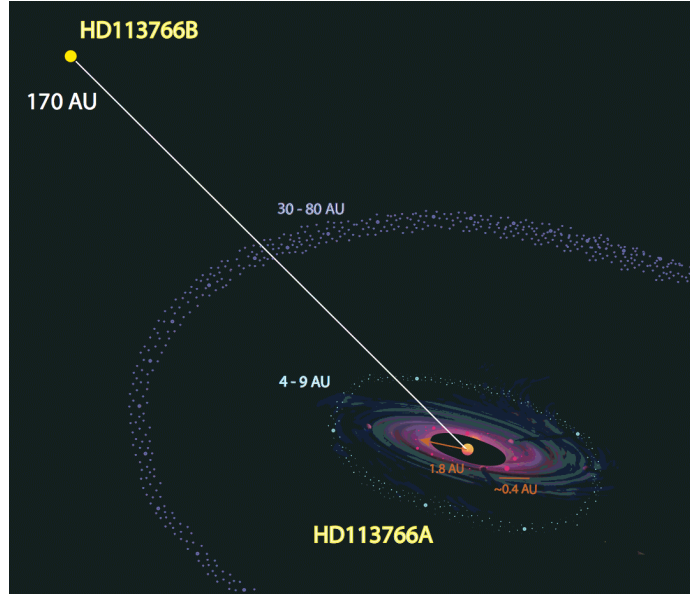

**Figure 6 — Schematic of the Binary Star System HD 113766.** The location of the S-asteroid-like dust causing the excess emission around HD 113766A found in this work, 1.8 ± 0.1 AU (1σ) AU (for $L_* = 4.4\ L_{solar}$, ), is consistent with but more tightly constrained than the range found by Meyer *et al.* 2001 (0.35–5.8 AU). The nearby binary component, HD 113766B, (distance not to scale), while potentially important for the dynamics of the system, is inconsequential for the energy balance of the inner-system dust causing the excess emission. The warm water-ice component detected can be anywhere within 9 AU of the HD 113766 stars, as determined by our analyses, including co-located within the dust ring, and another cold ice component is likely to be located at 30–80 AU.

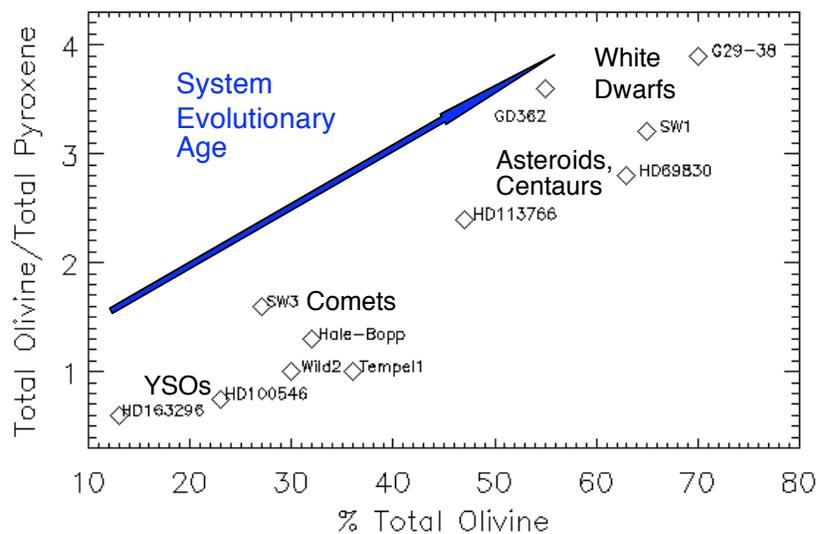

**Figure 7 — Silicate mineralogy for HD 113766** vs. that found in 4 comets systems (SW3, Sitko, *private communication* 2007; Hale-Bopp, Lisse *et al.* 2007a; Wild 2, Zolensky *et al.* 2007; and Tempel 1, Lisse, C. M. *et al.* 2006), the primitive YSO disk systems HD 100546 (Lisse, C. M. *et al.* 2007a) and HD 163296; the mature asteroidal debris belt system HD 69830 (dominated by P/D outer asteroid dust; Lisse, C. M. *et al.* 2007b); the Centaur SW-1 (Stansberry *et al.* 2004), and the ancient debris disk of white dwarfs G29-38 (Reach *et al.* 2005) and GD362 (Jura 2006). The general trend observed is that the relative pyroxene content is high for the most primitive material (i.e., YSOs), and low for the most processed (i.e., white dwarfs). HD 113766 plots to the primitive side of the asteroidal region.